\documentclass[preprint,12pt,authoryear]{elsarticle}

\usepackage{amssymb}
\usepackage{amsmath}
\usepackage{longtable}
\usepackage{subcaption}
\usepackage{endfloat}
\usepackage{rotating, graphicx}
\usepackage{multirow}

\usepackage[usenames]{color}
\usepackage{colortbl}

\usepackage{etoolbox}
\patchcmd{\pprintMaketitle}
{\hrule\vskip12pt}
{\hrule\vskip12pt\ifvoid\extrainfobox\else\unvbox\extrainfobox\par\vskip12pt\fi}
{}{}

\newenvironment{extrainfo}
{\global\setbox\extrainfobox=\vbox\bgroup\parindent=0pt }
{\egroup}
\newsavebox\extrainfobox

\begin{document}

\begin{frontmatter}



\title{Photometric Study of the C/2014 Q2 (Lovejoy) Comet Plasma Tails}


\author[a]{Valery V. Kleschonok\corref{cor1}}
\address[a]{Astronomical Observatory, Taras Shevchenko National University of Kyiv, Observatorna str. 3, Kyiv, 04053 Ukraine}
\cortext[cor1]{Corresponding author}
\ead{klev@observ.univ.kiev.ua}

\author[a]{Igor V. Luk'yanyk}

\author[b]{Yury M. Gorbanev}
\address[b]{Astronomical Observatory, Odessa National University, Shevchenko Park, 65014, Odessa, Ukraine}

\author[b]{Volodymyr I. Kashuba }

\begin{abstract}
Observations of C/2014 Q2 (Lovejoy) comet were carried out on 7th February, 2015, at the observation station in Mayaki village (No. 583 -- Odesa-Mayaki observatory). The integrated-light photometry of the comet was conducted using RC-800 telescope (D\,=\,80\,cm; F\,=\,214.0\,cm) with FLI MicroLine 9000 CCD camera. The photometric primary reductions included dark-frame subtraction and flat-field correction. The photometric study of the comet plasma tails was performed using an interactive program to construct a series of longitudinal and transverse profiles of individual tail rays. The Shulman diffusion model was applied to interpret the calculated photometric profiles. The comparison of the experimental profiles and those calculated theoretically from the diffusion model enabled us to estimate the following physical parameters of the comet plasma tail: acceleration a = 176\,m/sec$^2$ and lifetime of fluorescent ions $\tau$ = 2.7$\cdot$10$^3$\,sec; longitudinal and transverse diffusion coefficients, and magnetic flux density B = 97\,$\pm$\,5.3\,nT.   
\end{abstract}


\begin{extrainfo}
	\section*{Highlights}
	\begin{enumerate}
		\item[1] The Shulman-Nazarchuk diffusion model of cometary ion tail is used to determine of the physical parameters of the comet plasma tail.  
		\item[2] The magnetic flux density of comet C/2014 Q2 (Lovejoy) varied within the range of 79\,nT to 121\,nT with the mean value of 97\,$\pm$\,5.3\,nT.
		\item[3] A rather high mean value obtained is apparently due to the magnetic flux density increasing over time as the comet crossed the IMF sector boundaries quite a long time ago.
	\end{enumerate}
\end{extrainfo}

\end{frontmatter}


\section{Introduction}
Some comets develop separate tails of different types, including those of Type I as per the Bredichin classification of cometary tails. The study of the formation of Type I straight cometary tails contributed predominantly to the discovery of the solar wind. Due to anisotropic ejection of plasma from the near-nucleus coma, as well as plasma instabilities and solar wind inhomogeneities, plasma tails have filamentary structure in the form of tail rays. They are almost cylindrical (with a cross-section of (2$\div$3)$\cdot$10$^4$\,km diameter) with the ion concentration of about 10$^8$\,cm$^{-3}$. The angle (of several degrees) of aberration, i.e. deviation of the plasma tail axis from the prolonged radius vector, is correlated with the solar wind properties (such as solar wind velocity and interplanetary magnetic flux density) and comet’s orbital speed. A classification of plasma tail disturbances based on the survey of more than 500 photographs of the plasma tail of comet 1/P Halley has been suggested in \citet{Saito}. The main signatures in the plasma tail structure have been classified into outstanding streamers (S), outstanding rays (R), condensations (C), helices (H), arcades (A), kinks (K) and disconnection events. A disconnection event of the cometary plasma tail is one of the most spectacular phenomena observed in comets. Spectral analysis of the Type I tails has demonstrated that they are mainly comprised of ionised carbon monoxide CO$^+$. Fluorescence of the Type I tails is mostly attributable to the emission from the CO$^+$ ions which are optically active. Direct plasma measurements by space probes have shown the presence of other ions, namely N$_2^+$, OH$^+$, CO$_2^+$, H$_2$O$^+$, CN$^+$, CH$^+$, C$^+$ and O$^+$ in plasma tails. The above facts indicate that magnetic fields are involved in the formation of such cometary tails.\\ 
Hannes Alfv\'en was the first to suggest that the presence of interplanetary magnetic field in the Type I tails could be a quite reasonable explanation for some cometary events \citep{Alfven}. This hypothesis was further developed in later studies which often assumed that the solar-wind magnetic field maintains its regular structure despite undergoing some changes when interacting with a comet.\\ 
A contrary point of view has been presented in the study \citet{Dolginov}, according to which the cometary magnetic field can only be with random inhomogeneities. To analyze the plasma tail of the comet we can use the model of Wallis which is based on supposition that the tail can be considered as a jet structure that is imbedded in the solar wind. Mechanism which connects the solar wind and the tail plasma is considered as a plasma instability.\\
It should be noted that at the moment there is no completed quantitative model that can explain formation of the plasma tail and magnetosphere of the comet. When comparing the distribution of the surface brightness in head and tail of the certain comets with theoretically calculated within the most probable model, one can obtain quantitative estimates of a number of physical parameters which characterize physical properties of the plasma tail and interplanetary medium. To construct the theoretical model of the surface brightness distribution in a cometary plasma tail one can use several ways. The first way consists in the integration of the system of kinetic equations for the tail particles without taking into account their collisions, but with an account of their interaction with corpuscular solar radiation and self-consistent electric and magnetic fields. Small size of the comet's nucleus comparing with a tail size allows one to construct the models in a diffedrent way. Such a simplest model was published by \cite{Haser}, where the neutral gas density follows r$^{-2}$ profile with an additional exponential attenuation due to ionization losses. Another way is to use the Green function for instantaneous source. In this case the point source is a comet's nucleus with surrounding zone. Such models are suitable for observations and their interpretation. For instance, \citet{Goetz} use a simple 1D MHD model to interpret the magnetic field data during the entire mission ROSETTA. The model by \citet{Hansen} also gives an empirically determined gas production rate of comet 67P/Churyumov-Gerasimenko for the time from 2014 August to 2016 March 21. There are several limitations: (i) an assumption that the magnetic field in the solar wind is always perpendicular to the flow direction. (ii) MHD model is assumed, although the ion gyro-radius is of the order of the size of the interaction region. (iii) the model is only valid on the comet-Sun line.\\ 
Significant contribution in construction of this model was made due to the cosmic missions. In this case ROSETTA mission was the most importance since the observations were made on a quite long time interval. Direct measurements of the comet 21P/Giacobini-Zinner plasma tail 7,800\,km antisunward from the nucleus made by International Cometary Explorer probe are indicative of the presence of a well-developed magnetotail with magnetic flux density up to 60\,nT. The magnetic flux density in the tail of comet 67P/Churyumov-Gerasimenko estimated from recent measurements in the near-nucleus coma is about 100\,nT. Cometary tail fly-bys by spacecrafts are still a rare occurrence; thus, it is recommended to carry out observations using some other remote-sensing techniques which enable taking measurements regularly. Those could be photometric studies of cometary tails with application of a model of cometary tail formation which factors in the magnetic field. In this paper we used Shulman's model to interpret the ground based observations (see Sec. 2) of the comet C/2014 Q2 (Lovejoy).\\
C/2014 Q2 (Lovejoy) is a long-period comet discovered on 17 August 2014 by Terry Lovejoy. Its perihelion is 1.29 AU, eccentricity is 0.99811, orbital period is about 8000 years, inclination is 80.301$^\circ$. This comet has the prominent tail that is decisive for use the Shulman's model.\\\\

\section{The Shulman-Nazarchuk diffusion model}

As this model has not yet been published in international reviews or journals, we briefly describe it here. The diffusion model by G.K. Nazarchuk and L.M. Shulman \citep{Nazarchuk} is classified as a model algorithm. This model presents the plasma tail formation as the processes of diffusion and drift towards the comet tail. The model parameters can be determined by comparison of the longitudinal and transverse photometric profiles of the target comet plasma tail with the relevant profiles calculated theoretically. The obtained model parameters enable estimation of possible limits of acceleration and lifetime of fluorescent ions; longitudinal and transverse diffusion coefficients, as well as magnetic field of the plasma tail.\\
We assume that the surface density distribution in the cloud of molecules momentarily ejected from a point source of matter changes with time according to the following law:

\begin{equation}
\label{eq1} 
N = G(x,y,t),	
\end{equation}
where G(x,y,t) is the Green's function.\\
The laws of motion of the molecular cloud centre-of-mass in the source-related reference system have been taken as known:

\begin{equation}
\label{eq2}
\begin{cases}
x=\psi(t) 
\\
y=\phi(t),
\end{cases}
\end{equation}
where $\phi(t)$ and $\psi(t)$ are some functions of time. Under such conditions a continuous outflow of matter from the source (i.e. cometary nucleus) results in the formation of a tail (the plane of projection coincides with the orbital plane) for which the molecular surface density can be described by the formula:   
	
\begin{equation}
\label{eq3} 
n(x,y,t)=\int\limits_0^t f(t^*)G(x-\psi(t^*),y-\phi(t^*),t^*)dt^*.
\end{equation}
In this formula, time is counted back from the moment of observation t\,=\,0 to the past, i.e. f(t) is the source strength t units of time ago. A series of tail models can be obtained by selecting different functions f, G, $\psi$  and $\phi$ in the formula.\\
The following assumptions have been made in the diffusion model \citep{Nazarchuk}:
\begin{enumerate}
	\item[a)] sublimation of matter from the comet nucleus is steady (as the nucleus strength is constant) and started an infinite time ago
	
	\begin{equation}
	\label{eq4} 
	f(t)=const, t \rightarrow \infty;
	\end{equation} 
	 
	\item[b)] the centre-of-mass of each momentarily ejected packet of particles is accelerating a(t$^*$) at the same rate along the comet tail axis (t$^*$ is the age of the particle packet); thus, according to formula (\ref{eq2})
	
	\begin{equation}
	\label{eq5} 
	\phi(t^*)=0;  \psi(t^*)=\frac{a(t^*)^2}{2},
	\end{equation} 
\item[c)] cometary ions gain momentum when experiencing random impulses delivered by the self-consistent fields which fly through the comet tail. In other words, the process of interaction between cometary ions and the solar wind is assumed to be macroscopically stochastic; in this case, the motion of a cometary ion is the superposition of diffusion and drift towards the tail. Moreover, the distribution of particles in a packet which is expanding, obeys the normal probability law; at the same time, the dispersion (which varies along the coordinate axes) increases linearly with the packet’s age while the number of particles in the packet decreases exponentially. In such a case, G can be written as follows:
	
	\begin{equation}
	\label{eq6} 
	G=\frac{1}{4\pi}\frac{1}{t^*\sqrt{D_{||}^*D_{\bot}}}\exp{\bigg(-\frac{(x-a(t^*)/2)^2}{4D_{||}^*t^*} - \frac{y^2}{4D_{\bot}t^*} - \frac{t^*}{\tau}\bigg)},
	\end{equation} 
where D$_{||}^*$=D$_{||}$cos$^2$$\beta$+D$_{\bot}$sin$^2$$\beta$ is the diffusion coefficient in the projection plane along the tail axis; D$_{||}$, D$_{\bot}$ is the longitudinal and transverse diffusion coefficients; a is the acceleration; $\tau$ is the mean lifetime of fluorescent particles; t$^*$ is the age of a packet of particles; $\beta$ is the angle between the tail axis in space and the plane of projection. The x axis is directed along the tail axis (which points away from the Sun); the y axis is directed transversely. 
\end{enumerate}
Formula (\ref{eq6}) coincides with the Green's function for anisotropic diffusion of exponentially disappearing particles, hence the name of the model \citep{Nazarchuk}.\\ 
By substituting (\ref{eq6}) into formula (\ref{eq3}) we can obtain the surface density of particles which radiate at each point of the comet tail:

\begin{equation}
\label{eq7} 
n(x,y)=\frac{C}{4\pi\sqrt{D_{||}^*D_{\bot}}}\int\limits_0^\infty\exp{\bigg(-\frac{(x-at^2/2)^2}{4D_{||}^*t} - \frac{y^2}{4D_{\bot}t} - \frac{t}{\tau}\bigg)\frac{dt}{t}},
\end{equation} 
where C is some constant.
The following non-dimensional parameters have been set for the model: 
\begin{displaymath}
\Gamma=a \sqrt{\frac{\tau^{3}}{D_{||}^*cos\beta}},
X=\frac{x}{L_{||}}, Y=\frac{y}{L_{\bot}},
L_{||}=2\sqrt{D_{||}^*\tau}, L_{\bot}=2\sqrt{D_{\bot}\tau},
\Theta=\frac{t}{2\tau},
\end{displaymath}

\noindent where X and Y are the non-dimensional space coordinates among which X corresponds to the direction along the tail axis with the origin at the cometary nucleus while Y transverses the tail with the origin at the tail axis; $\Gamma$ is the non-dimensional acceleration parameter; $\Theta$ is a dimensionless time; L$_{||}$ and L$_{\bot}$ are the longitudinal and transverse dimensions; D$_{||}$ and D$_{\bot}$ are the longitudinal and transverse diffusion coefficients; D$_{||}^*$ is the diffusion coefficient in the plane of projection along the tail axis; $\beta$ is the angle between the tail axis in space and in the projection plane; $\tau$ is the mean lifetime of the particles; and a is the acceleration of ions in the tail.\\
The chemical composition of the fluorescent matter of the cometary tail is deemed to be constant. In this case, the surface brightness is proportional to the surface density of fluorescent particles: I(x,y)=kn(x,y), where k is the ratio coefficient  of surface brightness to surface density of particles on the image. Thus, the surface brightness can be written as follows:

\begin{equation}
\label{eq8} 
I(x,y)=k\frac{C}{4\pi\sqrt{D_{||}^*D_{\bot}}}\int\limits_0^\infty\exp{\bigg(-\frac{(X-\Gamma\Theta^2)^2 + Y^2}{\Theta} - \Theta\bigg)\frac{d\Theta}{\Theta}},
\end{equation}
According to \citet{Nazarchuk}, after setting the non-dimensional parameters in the diffusion model and using a logarithmic scale, the theoretical law of the surface brightness decrease is as follows: 

\begin{equation}
\label{eq9} 
-2.5\lg I = const - 2.5\lg \Phi(X,Y,\Gamma),
\end{equation}
where \begin{displaymath}
\Phi = \int\limits_0^\infty\exp{\bigg(-\frac{(X-\Gamma\Theta^2)^2 + Y^2}{\Theta} - \Theta\bigg)\frac{d\Theta}{\Theta}}.
\end{displaymath}
Assuming Y=0, we can obtain the longitudinal profile and, by fixing X, derive a set of transverse profiles. By adjusting model parameters $\Gamma$, L$_{||}$, a model curve that best matches the longitudinal profile calculated from the observations should be found. Then, having fixed the obtained best-fit parameters, we can determine the third parameter L$_{\bot}$ using the transverse photometric profile of the comet tail.\\
We assume that
\begin{equation}
\label{eq10} 
\frac{D_{\bot}}{D_{||}}=\frac{1}{1+\frac{\lambda^2}{r_i^2}},
\end{equation}
where $\lambda$ is the free path of electrons, and r$_i$ are the Larmour radii of ions. The free path of an electron along the magnetic field can be defined as the following ratio:
\begin{equation}
\label{eq11} 
\lambda \approx \frac{D_{||}}{\bar{v}_e},
\end{equation}
The Larmour radii of ions are equal to
\begin{equation}
\label{eq12} 
r_i = \frac{m_i \bar{v}_i}{eB},
\end{equation}
where $\bar{v}_i$ and$\bar{v}_e$  are the average thermal velocities of ions and electrons, which equal
\begin{displaymath}
\bar{v}_i = \sqrt{\frac{2kT}{m_i}}, \bar{v}_e = \sqrt{\frac{2kT}{m_e}},
\end{displaymath} 
and e is the elementary charge, $m_i$, $m_e$ are mass of ions and electrons and T is temperature.\\
By substituting expressions (\ref{eq11}) and (\ref{eq12}), as well as numerical values of the constants (given that CO$^+$ is the key fluorescent ion) into (\ref{eq10}), we can find a formula for estimation of the magnetic field in the comet tail:
\begin{equation}
\label{eq13} 
B \approx 2\times10^{11}\frac{\frac{T}{D_{||}}L_{||}}{L_{\bot}\cos\beta}[nT].
\end{equation}

\section{Observations and Data processing}
\begin{figure}[tp]
	\centering
	\includegraphics[width=1.0\linewidth]{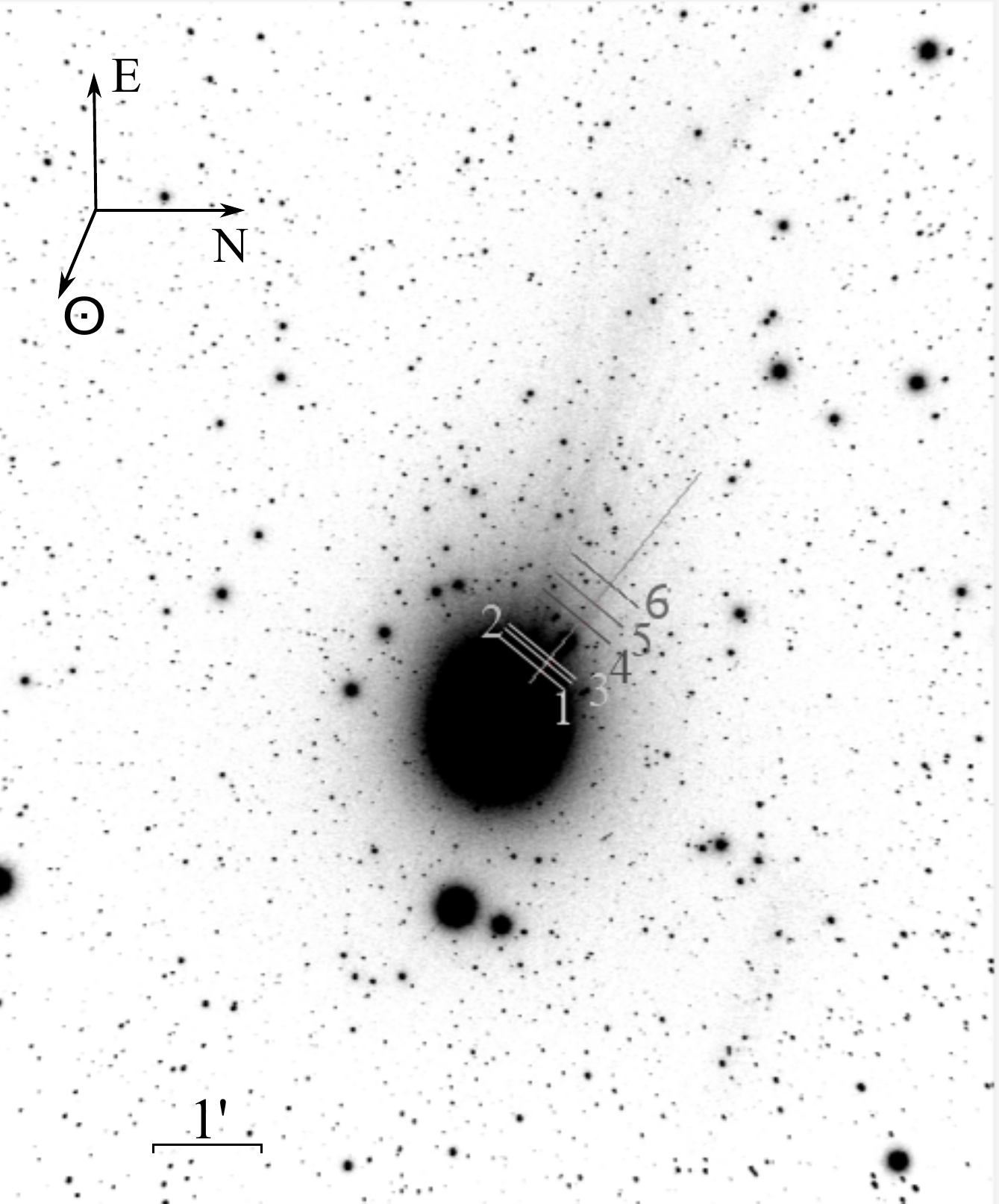}
	\caption{Observed image of C/2014 Q2 (Lovejoy) comet with a developed system of tail rays in February 7, 2015. Numbers indicate transverse profiles 1) at the distance of 1.356$\cdot$10$^5$\,km; 2) 1.494$\cdot$10$^5$\,km; 3) 1.607$\cdot$10$^5$\,km; 4) 2.417$\cdot$10$^5$\,km; and 5) 2.724$\cdot$10$^5$\,km.}
	\label{fig:fig1}
\end{figure}

Observations of C/2014 Q2 (Lovejoy) comet were carried out on the night 7$^{th}$--8$^{th}$ February, 2015, at the observation station in Mayaki  village \citep{Kashuba} (30.27$^\circ$\,E; 46.39$^\circ$\,N; 25\,m) using RC-800 telescope (D\,=\,800\,mm; F\,=\,2140\,mm; \citealp{Andrievsky}). At the time of observation, the geocentric distance to the comet $\Delta$\,=\,0.902\,AU; its distance from the Sun r\,=\,1.297\,AU; the phase angle $\alpha$\,=\,49.4$^\circ$; the ecliptic latitude $\beta$ of comet was about 20$^\circ$. Images of the comet were obtained using FLI MicroLine 9000 CCD camera with KAF-09000 sensor with a 3056$\times$3056 pixel array of 12$\times$12\,$\mu$m pixels. The scale in a frame is 1.16$^{\prime\prime}$/px. The size resolution for the faint stars is determined as approximately 2.5$^{\prime\prime}$. Several ions are taking part in the forming of the cometary tails, but the most prominent are only  CO$^+$  ions \citep{Wyckoff}. Therefore, it is better to use the CO$^+$ filter for observation. Nevertheless, we observed the comet  C/2014 Q2 (Lovejoy) without filters. The photometric primary reductions included dark-frame subtraction and flat-field correction. The field of view was about 1 degree thus allowing us to obtain images of the comet tail over a relatively long time interval (Fig. \ref{fig:fig1}). During the observations we observed the tail of a comet (separately) at some distance from the nucleus. On some frames it can be traced up to 1.8$^\circ$.\\

The preliminary processing of the comet images was performed using standard techniques (bias, dark, flat-field corrections). Next, the cometary tail images were processed using an interactive program in the MATLAB® environment (Mathworks product  https://www.mathworks.com). \\ 

Using this program enabled us to construct a series of cross-sections of the comet's Type I ion tail. To carry out processing, an individual ray not distorted due to overlapping by other rays was selected. To build the longitudinal and transverse profiles we used the ray which has the least number of stars at its image. Transverse profiles were constructed at those parts of the ray that were free of the stars. If some star was at the ray, such region of the ray was ignored. Also, the ray should be straight. It should not cross with other rays.	It should have the sharp boundaries in order to correctly take into account the background. The boundaries of the selected ray were determined using its image and further refined using the plotted cross-sections. As the comet tail was observed against the bright coma background, the coma's brightness was estimated on both sides of the tail on each cross-section. The background level in the middle of the tail image was calculated by linear interpolation of the background boundary values. The background level on both sides of the tail was estimated as the mean square value in a 3$\times$3 pixel square centered at the tail boundary for a given cross-section. In this way the applied technique that uses a background level along the sides of the ray, which allows to compensate the sky background and cometary coma too. 

\section{Data interpretation}
\begin{figure}[tp]
	\centering
	\includegraphics[width=1.0\linewidth]{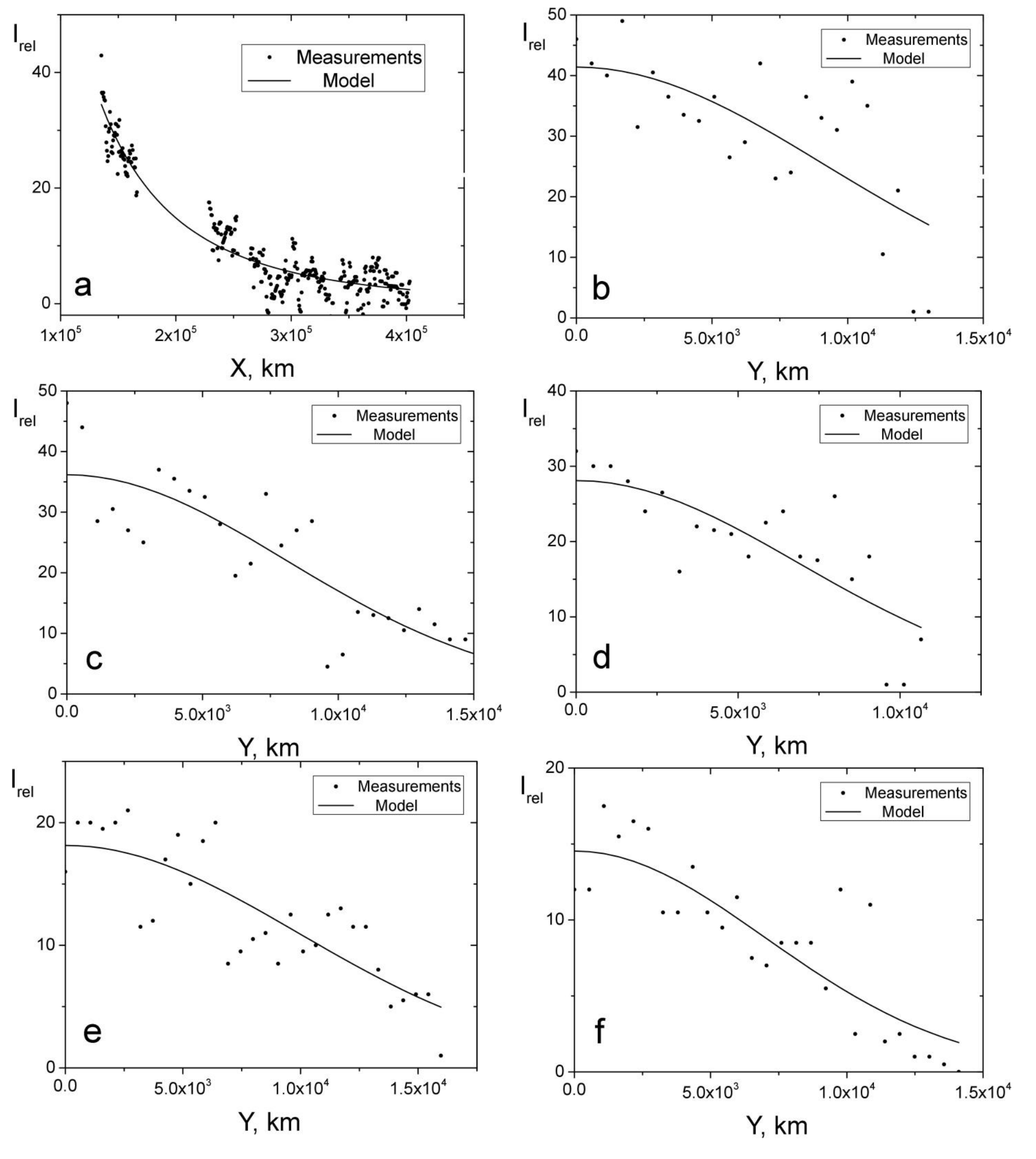}
	\caption{The longitudinal and cross-sectional dimensions of the C/2014 Q2 (Lovejoy) comet tail estimated with the Shulman model on 7$^{th}$ February, 2015: a) the longitudinal profile; b) the transverse profile at the distance of 1.356$\times$10$^5$ km; c) the transverse profile at the distance of 1.494$\times$10$^5$ km; d) the transverse profile 1.607$\times$10$^5$ km; e) the transverse profile at the distance of 2.417$\times$10$^5$ km; and f) the transverse profile at the distance of 2.724$\times$10$^5$ km. The flux is given in the relative values (I$_{rel}$)}
	\label{fig:fig2}
\end{figure}

In order to interpret the brightness distribution in the plasma tail within the diffusion model boundaries, the longitudinal dimensions L$_{||}$ and non-dimensional acceleration parameter $\Gamma$ were estimated by the longitudinal photometric profile while the cross-sectional dimensions L$_{\bot}$ were determined by the transverse profile of the tail. Places of selected profiles are shown in Fig. \ref{fig:fig1}. When estimating those parameters, the background level should be taken into account. The background was still markedly inhomogeneous despite all reductions as the tail was projected onto the coma and stellar field.\\
The transverse profiles were measured at different distances from the comet nucleus, hence several L$_{\bot}$dimensions determined. Figure \ref{fig:fig2} illustrates the results obtained. The distances presented in the caption to Figure \ref{fig:fig2} correlate with cuts 1,2,3,4,5 in Figure \ref{fig:fig1}.\\
After that the plasma acceleration in the comet tail should be estimated. Its precise determination is rather complicated and correlated to the solar wind parameters which are unknown. The limits of the ion acceleration are known from the observations of condensations in the cometary tails. The plasma tails are filled with structures such as bright knots or condensations which are basically ion concentrations. With the use of time sequence photographs, it is possible to follow these knots as a function of time for a period of a few hours to a day. From such measurements, the velocity and acceleration can be determined. Their velocities lie in the region of around 20 km/sec near the head to around 250 km/sec at distances far from the head. This corresponds to a value of (1 + $\mu$) of around 100 with wide variation \citep{Swamy}. Acceleration values can be expressed as absolute values as reported by \cite{Chernikov} or as the ratio of the plasma acceleration to the acceleration due to gravity of the Sun at a given heliocentric distance to the comet within the range of 100--1000 \citep{Daohan}. We have chosen an alternative way of expressing whereby the ion acceleration can be calculated using the following formula:
\begin{displaymath}
a=G\frac{M_{\odot}}{r^2}(1+\mu), 1 + \mu \geq 100,
\end{displaymath}
where G is the gravitational constant; M$_{\odot}$ if the mass of the Sun; r is the heliocentric distance of the comet. The mean value (1+$\mu$)\,=\,500 was used in our calculations.\\
The next step was to determine the average lifetime of ions with the formula as follows: 
\begin{displaymath}
\tau=\sqrt{\frac{\Gamma L_{||}}{2a\cos\beta}}.
\end{displaymath}
Then, the diffusion coefficients were determined:
\begin{displaymath}
D_{||}=\sqrt{\frac{aL_{||}^3}{8\Gamma\cos^3\beta}}\bigg(1-\bigg(\frac{L_{\bot}\sin\beta}{L_{||}}\bigg)^2\bigg);
\end{displaymath}
\begin{displaymath}
D_{\bot}=L_{\bot}^2\sqrt{\frac{a\cos\beta}{8\Gamma L_{||}}}
\end{displaymath}
The ion temperature in the cometary tail should be estimated in order to calculate the magnetic flux density (\ref{eq13}). Today, as a result of spacecraft probe missions, we understand the processes of the ion cometary tail formation much better. The cometary tail ions originate from two competing processes. The first process is the charge exchange of the solar-wind protons and ions with the neutral cometary molecules. This process produces high-energy ions (typically, a few hundred to a thousand eV) with a high characteristic velocity and velocity dispersion \citep{Combi}. Such ions are not able to form a narrow ray of the ion cometary tail. Another ion production process is photoionization. The low-energy or cold ions with temperatures of about 5--10\,eV (equivalent to 50,000--100,000\,K) are generated by photoionization \citep{Eriksson}. It is these ions which form ion tails observed in comets. Taking into account all the above, in all our calculations we used the mean ion temperature of 75,000\,K.\\
The calculations from the longitudinal profile gave the following results: L$_{||}$ = (2.1$\pm$0.2)$\times$10$^4$\,km; $\Gamma$ = 1.0$\pm$0.2. With the mean value (1+$\mu$)\,=\,500, the ion acceleration at the heliocentric distance of the comet a = 176\,сm/sec$^2$. The results of calculations using the cross-sectional profiles are given in the Table \ref{tabl:tab1}. The numbers in the first column correspond to the numbers of transverse profiles from Figure \ref{fig:fig1}. The column X of the table gives the distance from comet nucleus in kilometers of the transverse profile. The columns L$_{\bot}$ and $\Delta$L$_{\bot}$ of the table give transverse dimensions and its error in kilometers which are determined from spatial profile by Shulman-Nazarchuk model. The columns D$_{||}$ and D$_{\bot}$ of the table give diffusion coefficients in cm$^2$/sec. The errors of diffusion coefficient are not determined since they depend on ion acceleration, which is not precisely defined. Last column of the table give estimation of the magnetic field for each transverse profile in nT.
\begin{table}
	\caption{Results of determination of the magnetic field value within the Shulman-Nazarchuk model in C/2014 Q2 (Lovejoy) comet.}
	\label{tabl:tab1}
	\begin{tabular}{ccccccc}
		\hline\hline
		~ & ~ & ~ & ~ & ~ & ~ & ~ \\ 
		No & X, 10$^5$km & L$_{\bot}$,10$^3$ & $\Delta$L$_{\bot}$,10$^3$ & D$_{||}$,10$^{14}$ cm$^2$/sec & D$_{\bot}$,10$^{13}$ cm$^2$/sec & B,nT \\ 
		~ & ~ & ~ & ~ & ~ & ~ & ~ \\ 
		\hline
		~ & ~ & ~ & ~ & ~ & ~ & ~ \\
		1&1.356&8.4&1.2&5.8&6.5&79\\
		2&1.494&7.2&1.1&5.9&4.8&91\\
		3&1.607&6.0&1.0&6.0&3.3&108\\
		4&2.417&7.7&1.2&5.9&5.5&86\\
		5&2.724&5.3&1.1&6.0&2.6&121\\
		6&3.218&6.7&1.3&5.9&4.2&97\\
		\hline
	\end{tabular}
\end{table}

 Having analyzed the magnetic field estimates, assuming that the magnetic flux density in the plasma tail was zero at a certain instant of time before the observations ($\Delta$t\,=\,0), and having compared our findings with the results of the plasma tail study reported in \cite{Delva}, we suggest the following exponential law of the increasing magnetic flux density in the plasma tail newly formed after an earlier disconnection event:
\begin{displaymath}
B=B_f\bigg( 1-e^{-\Delta t/\tau^*} \bigg),
\end{displaymath}
where B$_f$ is the limiting magnetic flux density in the comet's plasma tail; $\Delta t$ is the period of time after the tail's disconnection event; $\tau^*$ is the time constant.\\
It can be deduced from the above formula and a rather high mean value of the magnetic flux density that the comet crossed the sector boundaries quite a long time ago. In the study \citep{Churyumov_Halley}, the following formula for determination of the time of the sector boundary crossing has been suggested:
\begin{displaymath}
t_0=\tau^*\ln\frac{B_2-B_1}{B_2e^{-t_1/\tau^*}-B_1e^{-t_2/\tau^*}}.
\end{displaymath}
Here t$_0$, t$_1$ and t$_2$ are the relevant instants of time in UTC. The moments t$_1$ and t$_2$ should be separated in time with an interval of at least 1\,--\,2 days. For this reason, unfortunately, we failed to determine the moment of the comet’s sector boundary crossing from our observations.

\section{Discussion}
\begin{figure}[tp]
	\centering
	\includegraphics[width=1.0\linewidth]{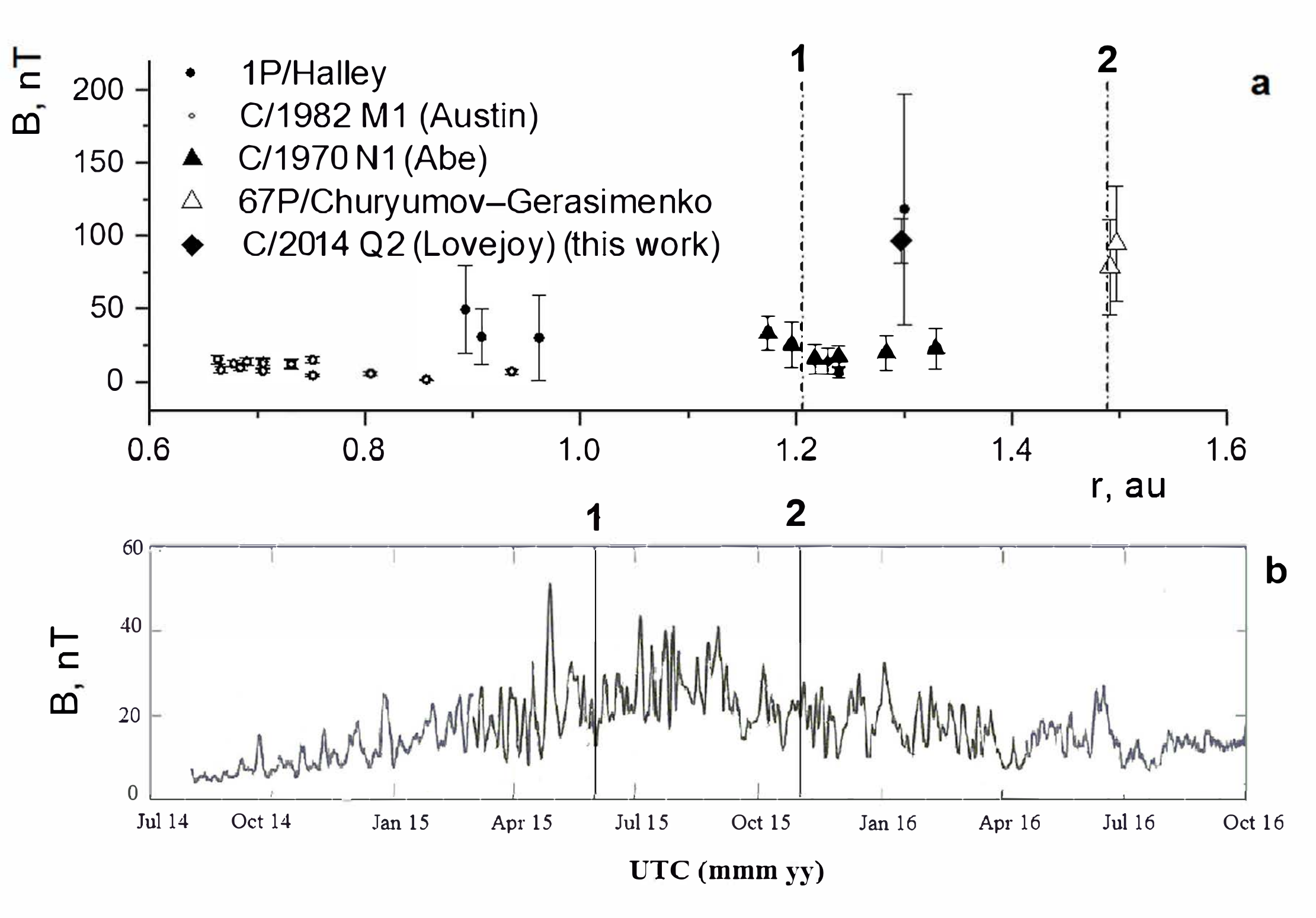}
	\caption{The magnetic field in the comet tails. a) -- the magnetic field in some comets (calculated within the Shulman model by different authors, see Discussion). b) -- the change in the magnetic field of comet 67P/Churyumov-Gerasimenko obtained through the ROSETTA mission for July 2014 to October 2016 (adapted from \citealp{Goetz}). The range of helicentric distances between vertical lines 1 and 2 on a panel a) corresponds to the same range of heliocentric distances of comet 67P/Churyumov-Gerasimenko marked on a panel b) by vertical lines 1 and 2.}
	\label{fig:fig3}
\end{figure}

Multiple effects which may impact the ion tail formation are not factored in the Shulman diffusion model; however, as this model is computationally simple and yields results which are in good agreement with the observation data, it can be employed to compare physical conditions in different comets. The Haser model of the coma gas production rates for certain emissions is a good example of such models. A'Hearn has reported that this model produces correct results even though its initial conditions carry no physical content \citep{AHearn}. For this reason, this model is widely used to find specific features of gas emission in different comets. The Shulman model has been tested during numerous studies of cometary plasma tails: comet 1P/Halley  \citep{Yan,Sizonenko_Halley,Churyumov_Halley}, C/1970 N1 (Abe) \citep{Shabas_Abe,Churyumov_Abe}, comet C/1982 Ml (Austin)  \citep{Shabas_Ostin}, comet C/1987 P1 (Bradfield) \citep{Borisenko}, 67P/Churyumov-Gerasimenko \citep{Churyumov_ChG,Shabas_ChG}, comet C/1976 R1 (West) \citep{Sizonenko_Vesta}.\\
Fig. \ref{fig:fig3} illustrates the results of calculations carried out using the Shulman model, as well as the data obtained by spacecraft missions. As can be seen, this model produces correct results.
The feasibility of using this model has been confirmed independently by the obtained lifetime of ions $\tau$. The ionized carbon monoxide CO$^+$ bands are the most prominent in comet-tail spectra. The spectroscopic observations of comet 29P/Schwassmann-Wachmann made by \cite{Cochran} have shown that the time scale of CO$^+$ ions creation is 1$\times$10$^5$\,sec at the heliocentric distance of 5.8\,AU. Assuming that the ion lifetime is $\sim$ r$^{-2}$, the estimated lifetime of ions at the heliocentric distance r\,=\,1.297\,AU is 5$\times$10$^3$\,sec. This value is close to the lifetime of 2.7$\times$10$^3$\,sec obtained using the Shulman model for the ion tail.\\
As is seen in Table \ref{tabl:tab1}, the magnetic flux density along the plasma tail varies widely around the mean value of 97\,$\pm$\,5.3\,nT.\\ 
In our opinion, such variations describe the precision of the method used, as well as spatial variations in the physical conditions of the solar wind. The accuracy of the magnetic flux density determination can be improved provided that the ion acceleration is estimated from the successive tail images by the shifts of inhomogeneities in the comet tail.

\section{Conclusions}
There are only a few models that allow to obtain from the observations specific physical parameters of a comet, for example, a magnetic flux density or acceleration and the lifetime of particles. In this article using the Shulman model we obtained that the magnetic flux density of comet C/2014 Q2 (Lovejoy) observed on the night 7$^{th}$\,--\,8$^{th}$ February, 2015, varied within the range of 79\,nT to 121\,nT with the mean value of 97\,$\pm$\,5.3\,nT. In our opinion, such variations describe the precision of the method used, as well as spatial variations in the physical conditions of the solar wind. The accuracy of the magnetic flux density determination can be improved provided that the ion acceleration is estimated from the successive tail images by the shifts of inhomogeneities in the comet tail. A rather high mean value obtained is apparently due to the magnetic flux density increase over a time as the comet crossed the IMF sector boundaries quite a long time ago. It should be noted that the values of the physical parameters that we have calculated are close to those obtained by the space mission of ROSETTA.\\
Also we want to note that the accuracy of the Shulman model strongly depends on the acceleration and temperature of cometary ions. Acceleration can be obtained from a sequence of observations of the emerging structures (by their displacement) at the tail of the comet. Ionic temperature can be obtained by assuming that narrow rays at the tail of the comet are formed by cold, low-energy ions with temperatures of about 5--10\,eV (equivalent to 50,000--100,000\,K). Such ions of comets are formed as a result of photoionization.
	
\section*{Acknowledgements}
V. Kleshchonok $\&$ I. Luk'yanyk research was supported through the project 16BF023-02 of the Taras Shevchenko National University of Kyiv.





\end{document}